\documentclass[12pt]{article}
\usepackage{graphics,epsfig}
\usepackage{amssymb}

\newcommand \be {\begin{equation}}
\newcommand \ee {\end{equation}}
\newcommand \ba {\begin{array}}
\newcommand \ea {\end{array}}
\newcommand \bea{\begin{eqnarray}}
\newcommand \eea{\end{eqnarray}}

\begin{document}

\begin{flushright}
\today
\end{flushright}

\vspace*{30mm}

\begin{center}
{\LARGE \bf DPEMC : A Monte-Carlo for Double Diffraction}

\par\vspace*{20mm}\par

{\large \bf M. Boonekamp $^{a, b}$, T. K\'ucs $^{b, c}$}

\bigskip

{\em $^a$ CE-Saclay,
F-91191 Gif-sur-Yvette Cedex, France} \\
{\em E-mail:} Maarten.Boonekamp@cern.ch \\
\vspace*{5mm}
{\em $^b$ CERN, CH-1211, Geneva 23, Switzerland} \\
\vspace*{5mm}
{\em $^c$ C.N.\ Yang Institute for Theoretical Physics,
Stony Brook University \\
Stony Brook, New York 11794-3840, U.S.A.} \\
{\em E-mail:} Tibor.Kucs@sunysb.edu

\end{center}
\vspace*{15mm}

\begin{abstract}

We extend the {\tt POMWIG} Monte Carlo generator developed by B.~Cox and J.~Forshaw,
to include new models of central production through inclusive and exclusive Double Pomeron Exchange
in proton-proton collisions. Double photon Exchange processes are described as well, both in proton-proton
and heavy-ion collisions. In all contexts, various models have been implemented, allowing for comparisons
and uncertainty evaluation and enabling detailed experimental simulations.

\end{abstract}

\newpage

\section*{Program summary}

\noindent {\it Title of the program:} {\tt DPEMC}, version {\tt 2.3}\\

\noindent {\it Computer:} any computer with the FORTRAN 77 compiler under the UNIX or Linux operating systems.\\

\noindent {\it Operating system:} UNIX; Linux versions 7.x, 8.x\\

\noindent {\it Programming language used:} FORTRAN 77\\

\noindent {\it High speed storage required:} $<$ 25 MB\\

\noindent {\it Keywords:} Proton-(anti)proton collisions, diffraction, double pomeron exchange, double photon exchange.\\

\noindent {\it Nature of the physical problem:} Proton diffraction at hadron colliders can manifest itself in many forms, and a variety of 
               models exist that attempt to describe it \cite{bialas,bpr,sci,cfpomwig,kmr}. This program implements some of the more 
               significant ones, enabling the simulation of central particle production through color singlet exchange between interacting 
               protons or antiprotons.\\

\noindent {\it Method of solution:} The Monte-Carlo method is used to simulate all elementary $2\rightarrow 2$ and $2\rightarrow 1$ processes available
               in {\tt HERWIG}. The color singlet exchages implemented in {\tt DPEMC} are implemented as functions reweighting the photon flux already 
               present in {\tt HERWIG}.\\

\noindent {\it Restriction on the complexity of the problem:} The program relying extensively on {\tt HERWIG}, the limitations are the same as in 
               \cite{herwig}.\\

\noindent {\it Typical running time:} Approximate times on a 800 MHz Pentium III : 5-20 minutes per 10000 unweighted events, depending on the
               process under consideration.

\newpage

\section{Introduction}

Doubly diffractive Higgs boson production has been much debated during the last decade, in a variety of theoretical approaches and
predictions \cite{bialas,bpr,sci,cfpomwig,kmr}. Now that the Tevatron is running, with both the D0 and CDF experiments well equipped for forward proton 
detection, there is good hope to clarify the situation in the mid-term. Meanwhile, potential upgrades of the LHC experiments 
ATLAS and CMS are being discussed, and a systematic evaluation of the DPE physics potential becomes necessary. The present work is 
meant as a step in this direction.

Available programs for diffraction simulation are {\tt POMPYT}
\cite{pompyt} and {\tt RAPGAP} \cite{rapgap}; the factorized model
of Cox and Forshaw, {\tt POMWIG} \cite{cfpomwig}; and the Soft 
Color Interaction model of Enberg, Ingelman and Timneanu \cite{sci}. {\tt POMWIG} is implemented as 
an extension to {\tt HERWIG} \cite{herwig} and relies on the following branchings:

\be
p \rightarrow p' + \mathrm{I\!P}, \mathrm{\,\,\,followed\,by\,\,\, } \mathrm{I\!P} \rightarrow q + X
\ee

\noindent where $p$ and $p'$ are the incoming and outgoing protons, the Pomeron is denoted $\mathrm{I\!P}$, and $q$ is the parton 
(quark or gluon) involved in the hard subprocess. In this $inclusive$ picture, the final state can be made in lowest order of e.g. jets, 
photon pairs, lepton pairs, or a Higgs boson, accompanied by remnants from the $\mathrm{I\!PI\!P}$ collision and the outgoing protons.

Technically, the simulation of $\gamma\gamma$ interactions in $ee$ collisions, provided by {\tt HERWIG}, is transformed into $pp$ induced 
$\mathrm{I\!PI\!P}$ interactions through a reweighting of the photon flux and a proper choice of the structure functions. The original 
implementation of this procedure, {\tt POMWIG}, was based on a Regge fit to the Pomeron flux and Pomeron parton distribution functions provided 
by the H1 experiment \cite{h1diffpdf}.

Note that nothing prevents us from applying the same machinery to simulate $exclusive$ $\mathrm{I\!PI\!P}$ interactions, i.e. processes 
where the entire $\mathrm{I\!PI\!P}$ center-of-mass energy is used in the production of a heavy central system. Most often, the Pomeron 
is seen as a gluonic object, so that the branchings

\be
p \rightarrow p' + \mathrm{I\!P}, \mathrm{\,\,\,followed\,by\,\,\, }
\mathrm{I\!P} \rightarrow g + g_{NP}
\ee

\noindent result in a hard gluon-gluon interaction, with e.g. jet
pairs, photon pairs or Higgs bosons in the final state. The color screening
soft gluon, $g_{NP}$, couples to its counterpart from the opposite beam. Pomeron 
remnants are absent, and the protons again leave intact. Contrarily to the inclusive processes, which are being measured at the 
Tevatron (through dijet production \cite{cdfdiffdijet}), the exclusive DPE has not yet been observed. This category of events is most appealing 
however, because Higgs boson production through this channel provides a very competitive measurement of its mass \cite{missingmass}, and 
possible determination of its spin and parity, which is a major challenge in non-diffractive channels.

It is also simple to describe $\gamma\gamma$ interactions in the same framework : this is just going back to the original {\tt HERWIG} 
where, to accommodate proton- or heavy ion-induced photon fluxes, we need to take into account the corresponding form factors. 
Compared to $ee$ induced $\gamma\gamma$ interactions, an additional complication comes from the fact that one needs to simultaneously 
forbid strong interactions between the nuclei. To achieve this, one generally works in impact parameter space and requires this impact 
parameter to be large enough for the nuclei not to geometrically overlap.
These interactions have interest in their own right (they are a basic ingredient of the study of ultra-peripheral heavy-ion collisions, 
for example), but also provide a natural lower bound on the exclusive central production in $pp$ collisions. In the worst case, these 
channels thus provide a conservative estimate of the physics potential of DPE.

This flexible framework allows to implement a number of models alternative to that of Cox and Forshaw in an analogous way. We focus 
here on the exclusive model of Bialas and Landshoff (BL) \cite{bialas}, and on its inclusive extension by Boonekamp, Peschanski and Royon 
(BPR) \cite{bpr}. On the $\gamma\gamma$ side, we propose an implementation of the computations by Cahn and Jackson (CJ) \cite{cahn}, Drees, Ellis 
and Zeppenfeld (DEZ) \cite{drees}, Papageorgiu \cite{papa}, and Budnev
et al \cite{budnev}.

We can unfortunately not pretend to be exhaustive, and the implementation of further models is left for future work. One notable example 
is the model of Khoze, Martin and Ryskin \cite{kmr}. We also escape discussing proton non-survival through soft parton rescattering; the 
cross-sections presented here should thus be understood as incomplete in that respect. Note however that the results of the inclusive 
DPE models can be rescaled to the measured DPE dijet cross-section, as in \cite{bpr}. On the other hand, in QED mediated diffraction, the 
proton collisions occur at much higher impact parameter, and the probability of parton rescattering is small \cite{kmr}.

The next section gives an overview of the models newly implemented (we leave most physical discussions aside and orient the reader to 
the references quoted above). Some results are then given in the third section, and the Appendix provides a guide to the package, and 
lists all settings and switches that control the output.

\section{DPE models implemented in {\tt DPEMC}}

This section describes the new models and processes which can be studied using {\tt DPEMC}. The model of Cox and 
Forshaw is described in \cite{cfpomwig} and of course still available.

\begin{figure} \center
\scalebox{0.98}[0.98]{\includegraphics*{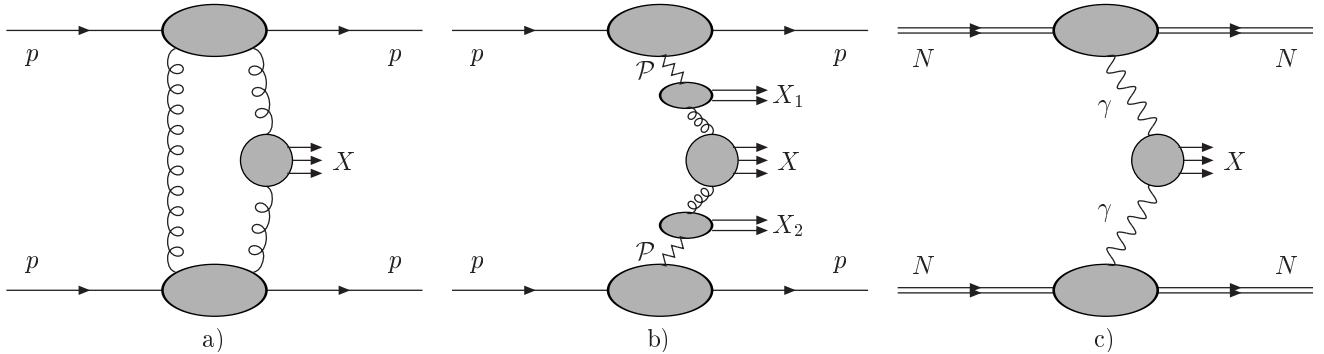}}
\caption{\label{dpefig} a) Exclusive QCD, b) Inclusive QCD and c) QED mediated diffractive reaction in $pp$ or Heavy-Ion collisions.} 
\end{figure}

\subsection{Exclusive DPE in the Bialas-Landshoff model \cite{bialas}}

The production cross section for a heavy system $X$ (such as a Higgs boson, a jet pair, or a photon pair) in exclusive DPE collisions:

\be \label{procExc}
p + p \rightarrow p + X + p,
\ee 

\noindent as illustrated in Fig.~\ref{dpefig}a, takes the form \cite{bialas,bpr}:

\be \label{sigmaExc}
d\sigma^{\rm (exc)} = \int d\xi_1 d\xi_2 \; F_{\mathrm{I\!P} / p}(\xi_1) F_{\mathrm{I\!P} / p}(\xi_2) 
\; d{\hat \sigma}(g g \rightarrow X).  
\ee

The convolution in Eq. (\ref{sigmaExc}) is over the momentum fractions of the initial protons, $\xi_1$ and $\xi_2$, carried away by the 
Pomerons. The flux of the Pomeron inside the proton is parametrized as:

\be \label{flux}
F_{\mathrm{I\!P} / p}(\xi) = {\cal N} \int {\rm d}v^2
\; \frac{e^{\beta \, v^2}}{{\xi}^{2 \alpha_\mathrm{I\!P} (v^2) - 1}}, 
\ee

where $v^2$ is related to the momentum transfer squared and the proton momentum loss through $v^2=t/(1-\xi)$. Here the parameters ${\cal N}^2 = 288/{\pi}^5$,
$\beta=4 \, {\rm GeV}^{-2}$ and the Regge trajectory 

\be
\alpha_\mathrm{I\!P}(t) = \alpha_\mathrm{I\!P}(0) + \alpha^{\prime}_\mathrm{I\!P} t,
\ee

\noindent with intercept $\alpha_\mathrm{I\!P}(0) = 1.08$ and slope $\alpha^{\prime}_\mathrm{I\!P} = 0.25 \, {\rm GeV}^{-2}$, are taken as 
in the original Bialas-Landshoff model, Ref. \cite{bialas}.

The partonic cross section $d\hat\sigma$ in Eq. (\ref{sigmaExc}) describes the possible hard subprocesses $gg \rightarrow X$, of which 
a list is given in Table~\ref{t1}. In this context, the cross-sections satisfy color and angular momentum selection rules;
namely the initial state is color singlet, and only the helicity configuration with $J_z = 0$ and $P = +1$ contributes to 
$\hat \sigma$ \cite{selrule}. These features have important consequences on the sensitivity to Higgs boson searches, since they imply important 
suppression of the backgrounds while being harmless to a naturally scalar and color-singlet signal.

\begin{table}
\begin{center}
\begin{tabular}{|l|l|} 
\hline
Central system & Partonic process \\
\hline\hline
Higgs bosons & $gg \rightarrow H$ \\
Dijets & $gg \rightarrow gg$, $gg \rightarrow q{\bar q}$\\
Photon pairs & $gg \rightarrow \gamma \gamma$ \\
\hline
\end{tabular}
\end{center}
\caption{Hard subprocesses relevant to exclusive DPE.}
\label{t1}
\end{table}

Note that properly speaking, the Bialas-Landshoff expressions have been derived for Higgs bosons and massive quarks only. We propose here
a generalization to gluon pair production, by using the corresponding color-singlet, $J_z=0$ subprocess cross-sections instead 
of the original ones.

\subsection{Inclusive DPE \cite{bpr}}

We implement the model of BPR as an alternative to that of Cox and Forshaw (CF) \cite{cfpomwig}. Although different in spirit, 
both lead to similar expressions and final states. The production cross-section for a system $X$ ($X$ = Higgs, Dijet, Diphoton, Dilepton) 
in inclusive DPE collision (see Fig.~\ref{dpefig}b):

\be \label{incdpe}
p + p \rightarrow p + X + p + \mathrm{Pomeron \, \, remnants},
\ee

\noindent can be expressed as follows \cite{bpr}:

\be \label{sigmaInc}
d\sigma^{\rm (inc)} = \sum_{i,j} \int dx_1 dx_2 \; d\xi_1 d\xi_2 \; F_{\mathrm{I\!P} / p}(\xi_1) F_{\mathrm{I\!P} / p}(\xi_2) \;  
f_{i/\mathrm{I\!P}}(x_1, \mu^2) f_{j/\mathrm{I\!P}}(x_2, \mu^2) \; d{\hat \sigma}(i j \rightarrow X).  
\ee

\noindent Besides the convolution over the proton momentum fractions, the integration runs over the 
Pomeron parton distribution functions. The Pomerons momentum fractions carried by the colliding partons are 
denoted $x_1$ and $x_2$. The proton induced Pomeron flux, $F_{\mathrm{I\!P} / p}(\xi_i)$, 
is the same as in the exclusive case (the normalization and the parameters $\beta$, $\alpha_\mathrm{I\!P}(0)$ and 
$\alpha^{\prime}_\mathrm{I\!P}$ are assumed identical to those used in \cite{bialas}).
The partonic content of the Pomeron is expressed in terms of distribution functions:

\be
f_{i/\mathrm{I\!P}}(x_i,\mu^2) = x_i \; G_{i/\mathrm{I\!P}}(x_i, \mu^2),
\ee

\noindent where $G_{i/\mathrm{I\!P}}(x_i, \mu^2)$ are the ``true densities'' of the partons inside the Pomeron as measured by the H1 
collaboration \cite{h1diffpdf}. They are evaluated at a scale $\mu^2$ given by the mass of the central system. The integral of 
$f_{i/\mathrm{I\!P}}(x_i,\mu^2)$ is normalized to 1, so that in the limit $f_{i/\mathrm{I\!P}}(x_i,\mu^2) \rightarrow \delta(x_i)$
the exclusive expressions are recovered.

Note that the cross-sections obtained in this way are rescaled in \cite{bpr} according to the double diffractive dijet cross-section
measured in \cite{cdfdiffdijet}; this is not done in the program.

Some of the possible hard subprocesses $gg, q\bar{q} \rightarrow X$ are listed in Table~\ref{t2}.

\begin{table}
\begin{center}
\begin{tabular}{|l|l|} 
\hline
Central system & Partonic process \\
\hline\hline
Higgs bosons & $gg, q{\bar q} \rightarrow H$ \\
Dijets & $gg \rightarrow gg ( q{\bar q})$, $qg \rightarrow qg$, $q{\bar q} \rightarrow gg (q{\bar q})$, 
$q q' \rightarrow q q'$ \\
Photon pairs & $gg \rightarrow \gamma \gamma$, $q {\bar q} \rightarrow \gamma \gamma$ \\
Lepton pairs & $q{\bar q} \rightarrow l {\bar l}$ \\
\hline
\end{tabular}
\end{center}
\caption{Hard subprocesses relevant to inclusive DPE.}
\label{t2}
\end{table}

\subsection{Two-photon reactions in $pp$ and heavy-ion collisions \cite{cahn,drees,papa,budnev}} 

The production cross section for a central system $X$ in heavy-ion collisions through double photon exchange (see Fig. \ref{dpefig}c):

\be \label{procHi}
N + N \rightarrow N + X + N, 
\ee 

\noindent can be approximated by a factorized form:

\be \label{sigmaHi}
d\sigma^{(\gamma \gamma)} = \int d\xi_1 d\xi_2 \; F_{{\gamma} / {\rm N}}(\xi_1) F_{{\gamma} / {\rm N}}(\xi_2) 
\; d{\hat \sigma}(\gamma \gamma \rightarrow X).  
\ee

\noindent The photon flux from an ion of electrical charge $Z$ and mass $M_N$ is parametrized as follows by DEZ \cite{drees}:

\be \label{fluxIon}
F_{{\gamma} / N}(\xi) = \frac{\alpha Z^2}{\pi} \, \frac{1}{\xi} \, 
\left[e^{-\xi^2 M_N^2/Q_0^2} + \left(1+\frac{\xi^2 M_N^2}{Q_0^2}\right){\rm Ei}(\xi^2 M_N^2/Q_0^2)\right],
\ee

\noindent where the momentum transfer scale $Q_0 \simeq 60 \, {\rm MeV}$ and ${\rm Ei}(x) = \int_x^{\infty} {\rm d}t \, e^{-t}/t$. 
Formula (\ref{fluxIon}) is valid for heavy nuclei only. An alternative formulation, derived in impact parameter space, yields
\cite{cahn,papa}:

\be \label{fluxFac}
F_{\gamma/N} (\xi,\xi_0) = \frac{2 \alpha Z^2}{\pi \xi} 
\left[\frac{\xi}{\xi_0} K_0\left(\frac{\xi}{\xi_0}\right) K_1\left(\frac{\xi}{\xi_0}\right) - 
\frac{1}{2}\left(\frac{\xi}{\xi_0}\right)^2\left(K^2_1\left(\frac{\xi}{\xi_0}\right)-
K^2_0\left(\frac{\xi}{\xi_0}\right)\right)\right].     
\ee

\noindent In Eq.~(\ref{fluxFac}), $K_{0,1}(\xi/\xi_0)$ are the Bessel functions, while $\xi_0 \equiv 1/(R M_N)$ with $R$ the radius 
of the ion. As discussed in the Introduction, this parameter appears because the impact parameter space integration is limited to the 
region where the colliding nuclei do not geometrically overlap. One can use $R \sim 1.2 A^{1/3}$ for collisions of ions of atomic number 
$A$, or the proton radius to estimate the $\gamma\gamma$ cross-sections in proton-proton collisions. The partonic cross sections relevant 
in this context are listed in Table~\ref{t3}.

\begin{table}
\begin{center}
\begin{tabular}{|l|l|} \hline
Central system & Partonic process \\
\hline\hline
Higgs & $\gamma \gamma \rightarrow H$ \\
Dijet & $\gamma \gamma \rightarrow q{\bar q}$ \\
Lepton pairs & $\gamma \gamma \rightarrow l {\bar l}$ \\
Photon pairs & $\gamma \gamma \rightarrow \gamma \gamma$ \\
\hline
\end{tabular}
\end{center}
\caption{Hard subprocesses occurring in double photon exchange.}
\label{t3}
\end{table}

\section{Results and Perspectives}

\begin{figure}[!ht]
\begin{center}
\begin{tabular}{cc}
\epsfig{file=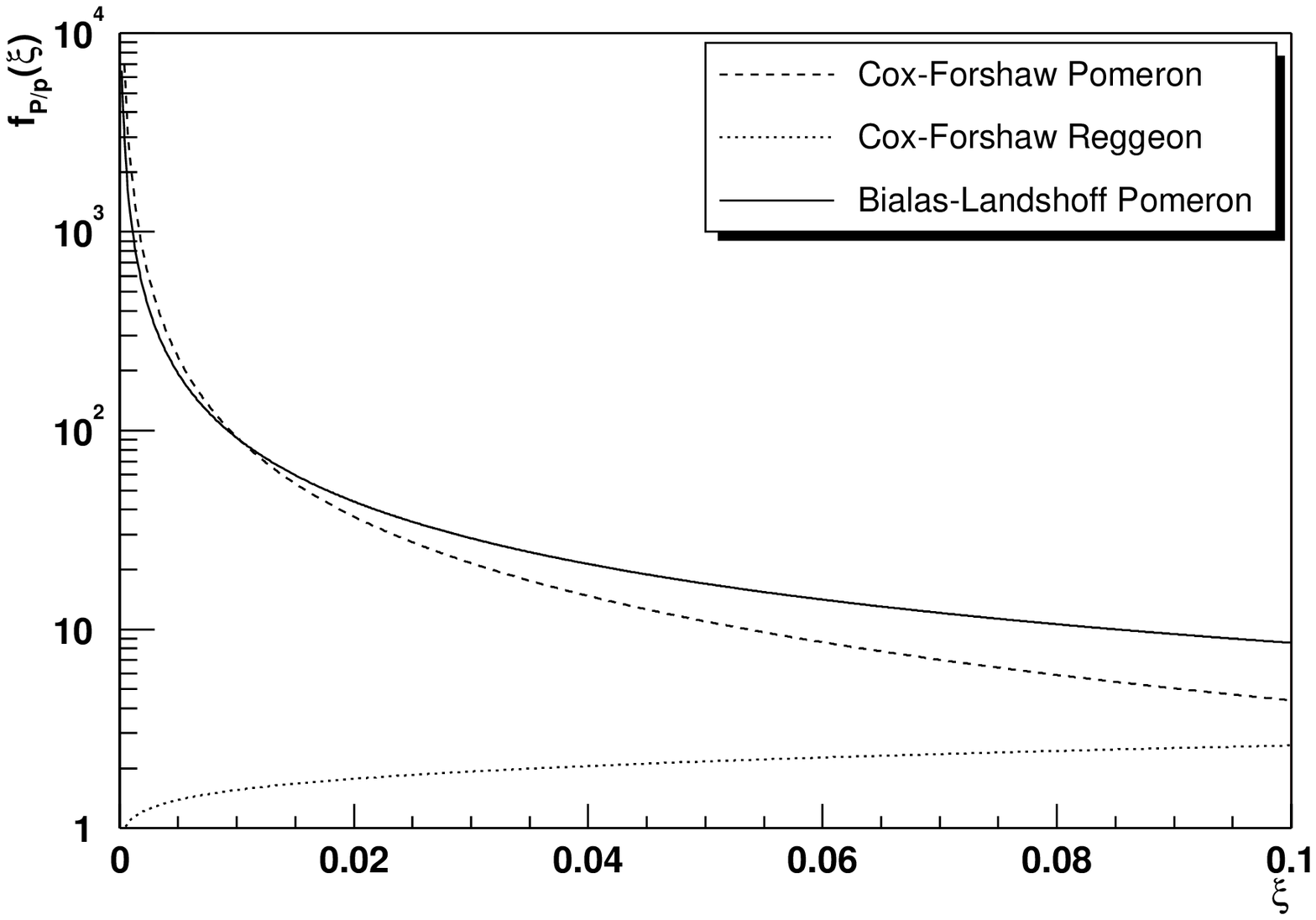,width=0.47\textwidth} &
\epsfig{file=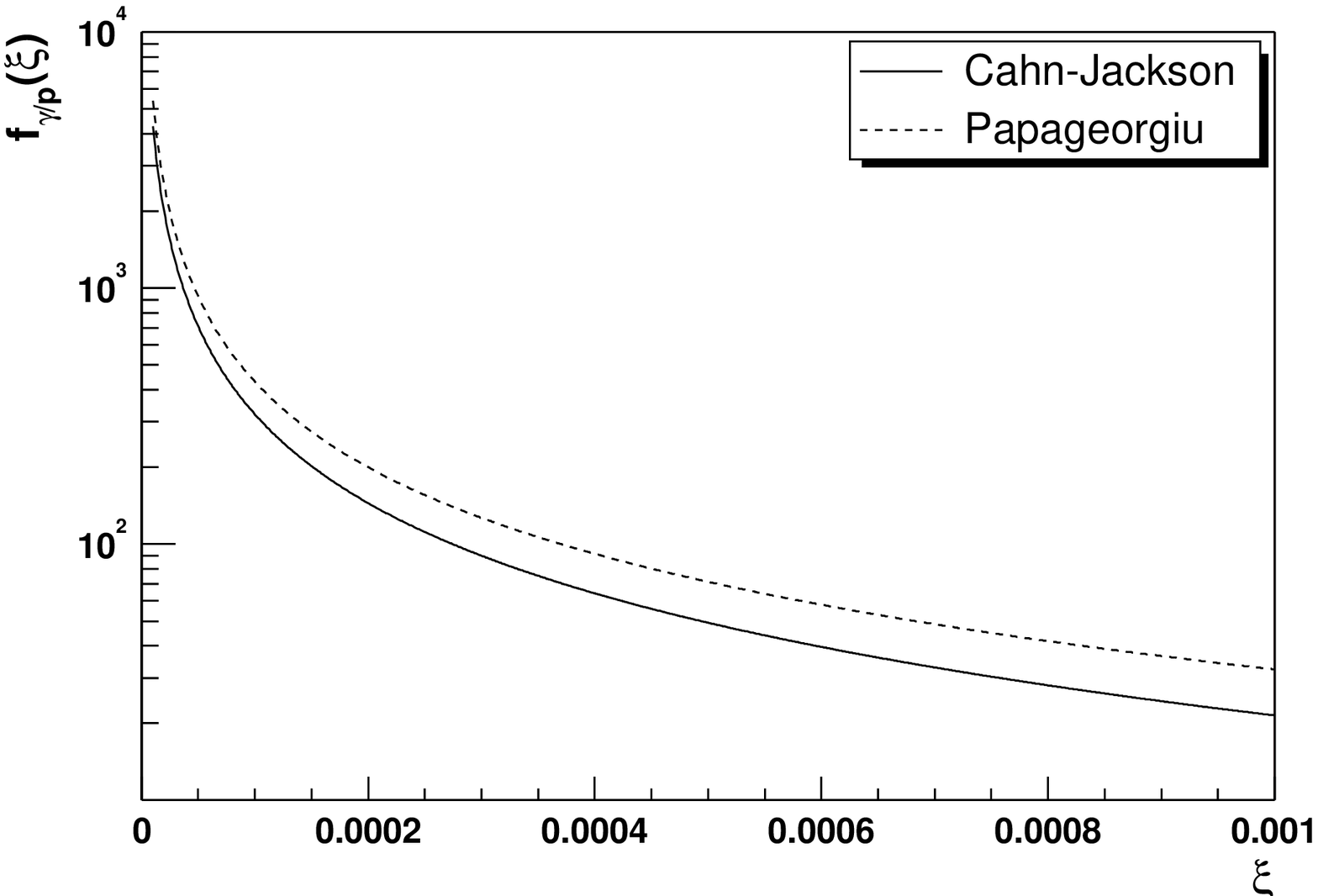,width=0.47\textwidth} \\
\epsfig{file=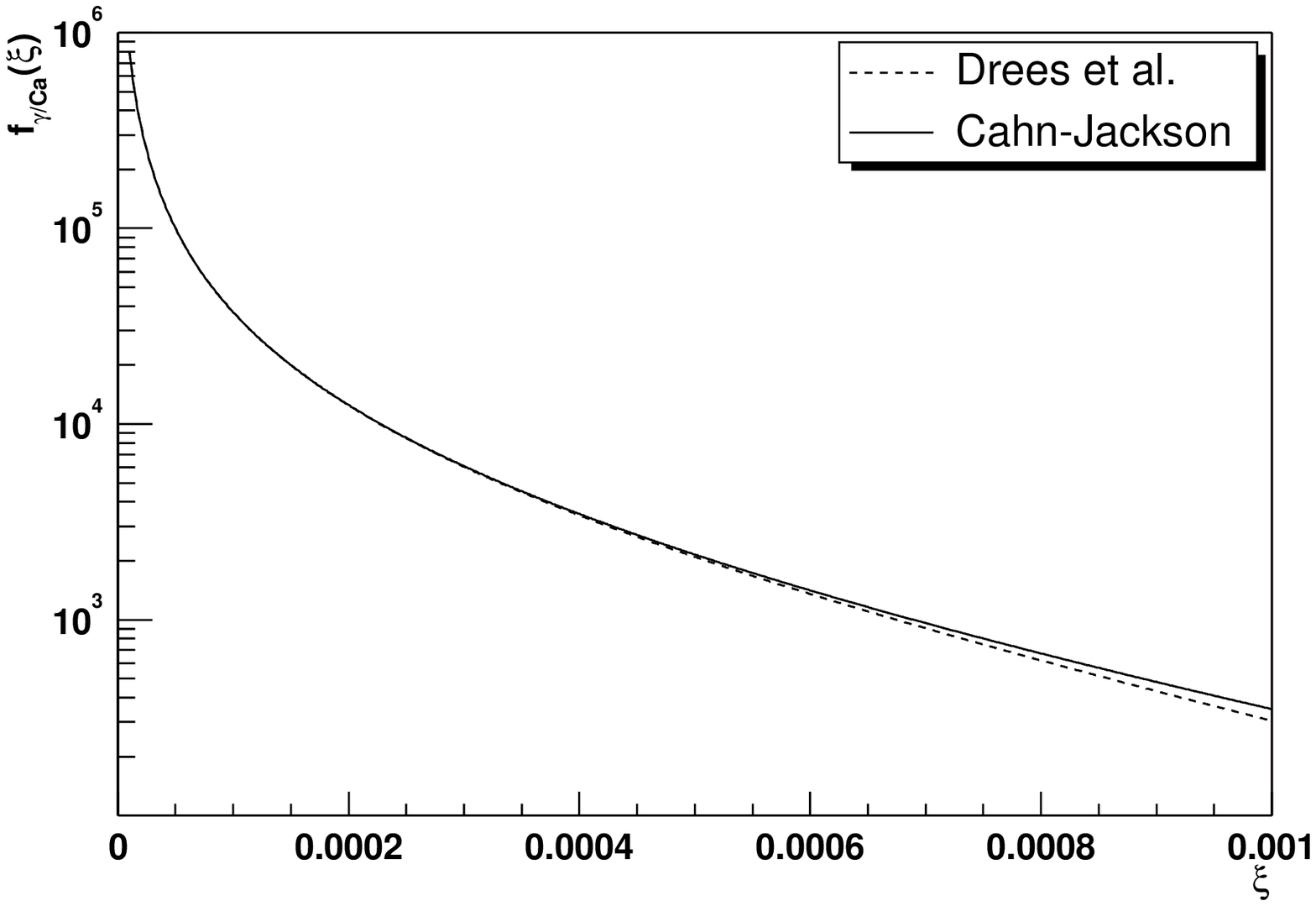,width=0.47\textwidth} &
\epsfig{file=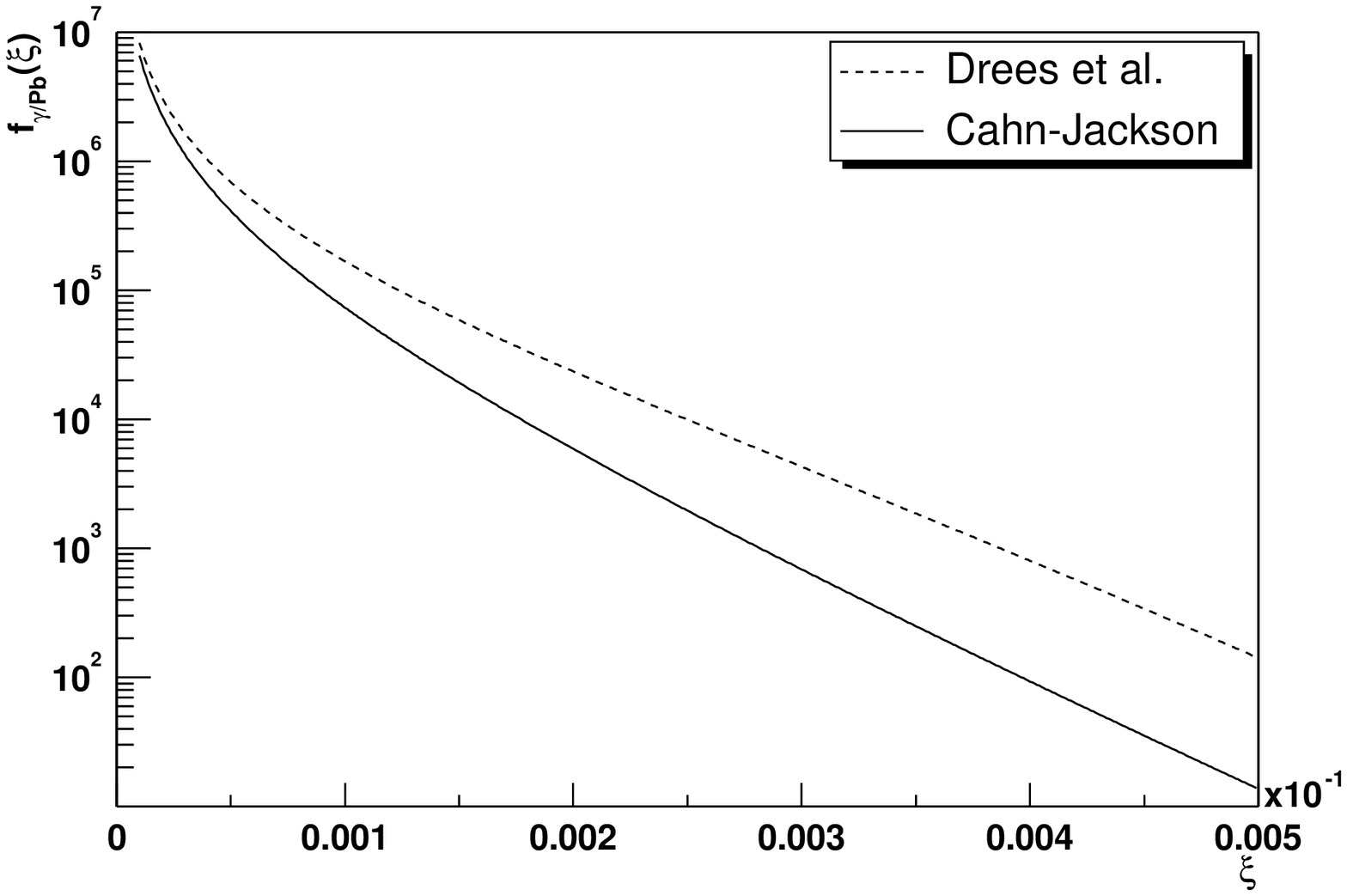,width=0.47\textwidth} \\
\end{tabular}
\end{center} 
\caption{Beam induced fluxes implemented in the program. {\it Upper left} : Pomeron flux according to \cite{cfpomwig} and \cite{bialas,bpr}, 
and the Reggeon flux of \cite{cfpomwig}. {\it Upper right} : proton induced photon flux according to \cite{cahn} and \cite{papa}. 
{\it Lower left and right} : Calcium and Lead induced photon flux following \cite{cahn} and \cite{drees}.}
\label{fig1}
\end{figure}

All implemented fluxes are displayed in Figure~\ref{fig1}. Apart from normalization, the CF and BL Pomeron fluxes differ mainly because
of different choices of the Pomeron intercepts ($\alpha_\mathrm{I\!P}(0)=1.2$ is taken by CF, while $\alpha_\mathrm{I\!P}(0)=1.08$ is used 
by BL and BPR). The nucleus induced photon flux is illustrated for the proton ($f_{\gamma/p}$), Calcium ($f_{\gamma/Ca}$), and Lead ($f_{\gamma/Pb}$).

\begin{figure}[!ht]
\begin{center}
\epsfig{file=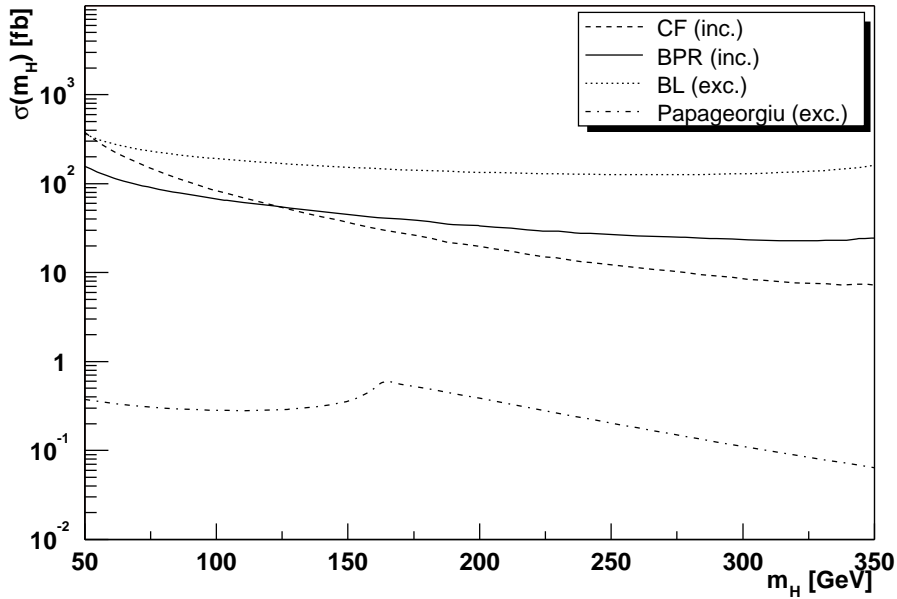,width=0.75\textwidth}
\end{center}
\caption{Inclusive Higgs boson production cross-sections following CF  \cite{cfpomwig} and BPR \cite{bpr}.
The exclusive predictions of BL \cite{bialas} and Papageorgiu \cite{papa} are also illustrated.}
\label{fig2}
\end{figure}

\begin{figure}[!ht]
\begin{center}
\epsfig{file=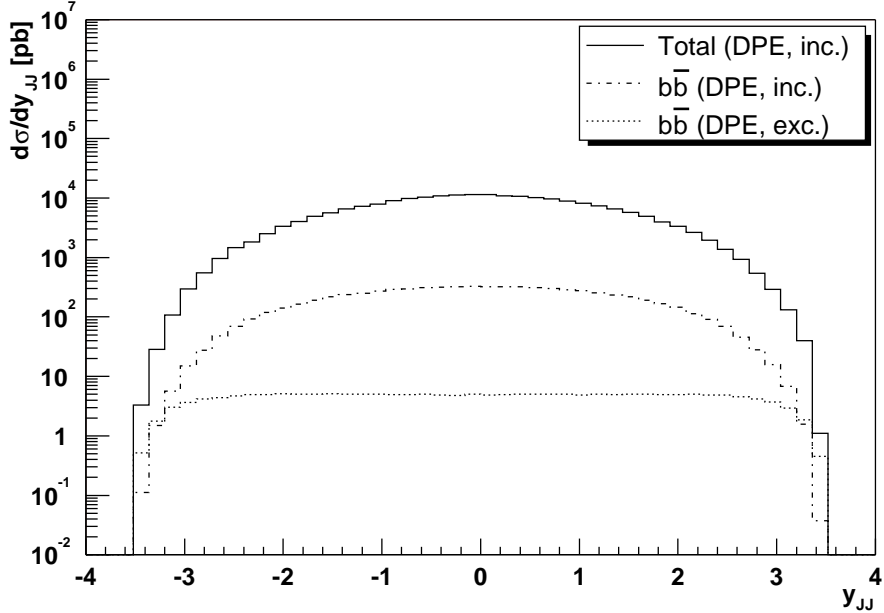,width=0.75\textwidth}
\end{center}
\caption{Distributions of the average dijet rapidity ($y_{JJ}=(y_1 + y_2)/2$) in the inclusive model of BPR \cite{bpr},
compared to the exclusive model of BL \cite{bialas}. The total inclusive dijet rate is displayed, as well as its $\mathrm{b\bar{b}}$
component, to be compared to the exclusive $\mathrm{b\bar{b}}$ cross-section.}
\label{fig3}
\end{figure}

Figure~\ref{fig2} illustrates Higgs boson production cross-sections in various models, namely the inclusive models of CF and BPR, the exclusive 
cross-section by BL, and the photon-mediated cross-section by Papageorgiu. Figure~\ref{fig3} displays the ``background'' in the BPR model (the 
total and $\mathrm{b\bar{b}}$ cross-sections are displayed) and in the exclusive BL model.  The exclusive $\mathrm{b\bar{b}}$ prediction is 
much suppressed as expected, and is more distributed towards small angles than its inclusive counterpart.

Considering the above examples and others that can be worked in an analogous way, it is possible using this program to perform complete studies 
of experimental sensitivity to diffractively produced Higgs bosons. These studies can be done following models of inclusive or exclusive double 
Pomeron exchange, and of double photon exchange in proton or heavy ion
collisions. Supersymmetric enhancements of the quark and gluon couplings to the Higgs boson can also be included as in {\tt HERWIG}.

The implementation of further models of hard diffraction, such as that
of Ref.~\cite{kmr}, will follow in subsequent versions of the
program. Another important addition will be the inclusion of
diffractive proton dissociation. Finally, since the beam particles
currently need to be defined as electrons, it is impossible to simulate soft
underlying events or multiple parton interactions. This shortcoming
will also be handled in the near future.

In summary, the aim of this work is not to promote one model or the other, but to provide a tool that should allow to determine those 
that are compatible with the available and forthcoming data, and to prepare the analyses and constrain the predictions for the LHC. 

\section{Acknowledgments}
We wish to thank B. Cox, J. Forshaw, V.A. Khoze, Ch. R. Peschanski, 
M. Rijssenbeek, C. Royon, M.H. Seymour and B.R. Webber for many helpful discussions.
One of us (T.K.) is grateful to M.~Rijssenbeek and CERN for support. 

\appendix

\section{Manual} \label{appb}

The source code for the program can be obtained from the authors or from Ref.~\cite{dpemccode}, and includes an example main program.
Some aspects of the implementation of {\tt DPEMC} are described in Appendix \ref{appa}; here we discuss 
how to run the program. The processes are controlled using the following flags:

\begin{itemize}
\item {\tt IPROC,ID,IL} select the hard process as in {\tt HERWIG};
\item {\tt TYPEPR} controls how the hard cross-section is to be evaluated; inclusively ({\tt TYPEPR='INC'}) or only the color-singlet $J_Z=0$ 
amplitude ({\tt TYPEPR='EXC'});
\item {\tt TYPINT} switches between photon initiated processes ({\tt TYPINT='QED'}), and gluon/quark initiated processes ({\tt TYPINT='QCD'});
\item \texttt{NSTRU} controls the $\mathrm{I\!P}$ or $\gamma$ flux; it can take values from 9 to 15, with meanings detailed in Table \ref{t4}.
For Pomeron fluxes, it controls the parton density functions to be used as well;
\item when {\tt NSTRU=11}, {\tt GAPSPR} and {\tt CDFFAC} set the absolute cross-section normalizations. {\tt GAPSPR} controls the gap survival 
probability for exclusive DPE processes ({\tt TYPEPR='EXC'}); typical values are 0.1 at the Tevatron, and 0.03 at the LHC \cite{kmr}. 
{\tt CDFFAC} controls the normalization of inclusive DPE processes ({\tt TYPEPR='INC'}), by comparison of the raw dijet computation to the 
measurement of \cite{cdfdiffdijet}; the value 3.8 was found in \cite{bpr}.
\item when appropriate, {\tt AION} and {\tt ZION} set the atomic and proton number of the colliding nuclei.
\end{itemize}

\begin{table}
\begin{center}
\begin{tabular}{|l|l|} \hline
{\tt NSTRU} & Flux and structure functions \\
\hline\hline
 9 & Factorized model, Pomeron flux \cite{cfpomwig} \\
10 & Factorized model, Reggeon flux \cite{cfpomwig} \\
11 & Bialas-Landshoff Pomeron flux \cite{bialas,bpr} \\
12 & QED flux from Cahn, Jackson \cite{cahn}; $R \sim 1.2 A^{1/3}$ \\
13 & QED flux from Drees et al., valid for heavy ions only \cite{drees} \\
14 & QED flux in $pp$ collisions, from Papageorgiu \cite{papa} \\
15 & QED flux in $pp$ collisions, from Budnev et al \cite{budnev} \\
\hline
\end{tabular}
\end{center}
\caption{Possible values of {\tt NSTRU} and their meaning.}
\label{t4}
\end{table}

As in {\tt POMWIG}, the beam particles need to be set as electrons in the main program, and the {\tt HERWIG} facility 
generating $\gamma\gamma$ interactions is used and reweighted to the user's needs, depending on the above settings.

Since {\tt NSTRU} controls which flux is to be used in the cross-section evaluations, the introduction of {\tt TYPEPR} and {\tt TYPINT} may 
seem somewhat redundant. This is however needed for technical reasons. First, exclusive $\mathrm{I\!PI\!P}$ interactions can not be treated
as such since the gluon initiating the hard process carries the full momentum of its parent ``beam particle'', provoking errors in {\tt HERWIG}.
To circumvent this problem, we needed to modify a few {\tt HERWIG} routines. For example, the routine evaluating $\gamma\gamma \rightarrow 
f\bar{f}$ has been modified so that it can compute the color-singlet, $J_Z=0$ \, $gg \rightarrow q\bar{q},gg$ cross-section, depending on the 
values of {\tt TYPINT} and {\tt TYPEPR}. 

For example, to simulate $b\bar{b}$ production in the exclusive Bialas-Landshoff model, one thus needs to request $\gamma\gamma \rightarrow b
\bar{b}$ via {\tt IPROC=16005}, set {\tt TYPINT='QCD'}, {\tt TYPEPR='EXC'} so that the color-singlet, $J_Z=0$ \, $gg \rightarrow q\bar{q}$ 
cross-section is actually computed, and {\tt NSTRU=11} requesting the Bialas-Landshoff flux. An exhaustive list of consistent flag settings is
given in Table \ref{t5}. These settings need to be followed strictly,
the program is producing unpredictable results otherwise\footnote{We
have tried to make the program detect unsound settings as much as possible.}.

It was also needed to provide some additional routines, for the computation of the $\gamma\gamma \rightarrow H$ and color-singlet, $J_Z=0$ \,
$gg \rightarrow \gamma\gamma$ cross-sections. For completeness, a computation of $\gamma\gamma \rightarrow \gamma\gamma$ is also provided. A list
of the new {\tt IPROC} values, and of {\tt ID} settings specific to {\tt DPEMC} is given in Table~\ref{t7}.

Finally, we provide an additional routine that corrects the event record after each event is finalized (remember that all events are recorded 
as $ee$ induced $\gamma\gamma$ interactions). This is needed for readability, but also to enable correct experimental simulations. When called, 
this routine resets the beam particles as protons, the secondary photon ``beams'' can be redefined as Pomerons or Reggeons, depending on the value 
of {\tt NSTRU}, and the particles initiating the hard process, always stored as photons when {\tt TYPEPR='EXC'}, are redefined as gluons when 
{\tt TYPINT='QCD'}.

\begin{table}
\begin{center}
\begin{tabular}{|llccc|} \hline
\multicolumn{5}{|c|}{}\\
Final state & \texttt{IPROC} & \texttt{TYPINT}/\texttt{TYPEPR} & {\tt ZION/AION} & \texttt{NSTRU} \\
\multicolumn{5}{|c|}{}\\
\hline\hline
\multicolumn{5}{|c|}{}\\
\multicolumn{5}{|c|}{Inclusive DPE:}\\
\multicolumn{5}{|c|}{}\\
Higgs bosons & 1600+\texttt{ID} & {\tt QCD/INC} &  - & 9, 10, 11 \\
Dijets       & 1500             & {\tt QCD/INC} &  - & 9, 10, 11 \\
Lepton pairs & 1350+\texttt{IL} & {\tt QCD/INC} &  - & 9, 10, 11 \\
Photon pairs & 2200             & {\tt QCD/INC} &  - & 9, 10, 11 \\
\multicolumn{5}{|c|}{}\\
\hline
\multicolumn{5}{|c|}{}\\
\multicolumn{5}{|c|}{Exclusive DPE:}\\
\multicolumn{5}{|c|}{}\\
Higgs bosons & 9900+\texttt{ID} & {\tt QCD/EXC} & - & 11 \\
Dijets       & 6000+\texttt{ID} & {\tt QCD/EXC} & - & 11 \\
Photon pairs & 9800             & {\tt QCD/EXC} & - & 11 \\
\multicolumn{5}{|c|}{}\\
\hline
\multicolumn{5}{|c|}{}\\
\multicolumn{5}{|c|}{Double photon exchange:}\\
\multicolumn{5}{|c|}{}\\
Higgs bosons & 9900+\texttt{ID} & {\tt QED/EXC} &  Z/A ; 1/1 & 12 or 13 ; 14\\
Dijets       & 6000+\texttt{ID} & {\tt QED/EXC} &  Z/A ; 1/1 & 12 or 13 ; 14\\
Lepton pairs & 6006+\texttt{IL} & {\tt QED/EXC} &  Z/A ; 1/1 & 12 or 13 ; 14\\
Photon pairs & 9800             & {\tt QED/EXC} &  Z/A ; 1/1 & 12 or 13 ; 14\\
\multicolumn{5}{|c|}{}\\
\hline
\end{tabular}
\end{center}
\caption{Processes available for simulation, and the corresponding settings. The values of {\tt IPROC}, {\tt ID} and {\tt IL} are set as in 
HERWIG \cite{herwig}, except for the new settings detailed in the Table~\ref{t7}. The flags should be set in the main program, of which
examples can be obtained with the source code.}.
\label{t5}
\end{table}

\begin{table}
\begin{center}
\begin{tabular}{|lll|} \hline
\multicolumn{3}{|c|}{}\\
{\tt IPROC} & {\tt ID}/{\tt IL} & Process\\
\multicolumn{3}{|c|}{}\\
\hline\hline
\multicolumn{3}{|c|}{}\\
1600+{\tt ID} & {\tt ID}=1,...,6   & $H \rightarrow q\bar{q}$ (resp. d,u,s,c,b,t)\\
              & {\tt ID}=7,8,9   & $ H \rightarrow l^+l^-$ (resp. $e^+e^-$,$\mu^+\mu^-$,$\tau^+\tau^-$)\\
              & {\tt ID}=10,11 & $H \rightarrow$ $W^+W^-$, $ZZ$\\
              & {\tt ID}=99    & All decay modes\\
\multicolumn{3}{|c|}{}\\
1350+{\tt IL} & {\tt IL}=0     & All leptons (incl. neutrinos)\\
              & {\tt IL}=1,...,6   & resp. e,$\nu_e$,$\mu$,$\nu_\mu$,$\tau$,$\nu_\tau$\\
\multicolumn{3}{|c|}{}\\
\hline
\end{tabular}
\end{center}
\caption{Reminder of the {\tt ID,IL} settings for Higgs boson and lepton pair production in {\tt HERWIG}.}
\label{t6}
\end{table}

\begin{table}
\begin{center}
\begin{tabular}{|llcl|} \hline
\multicolumn{4}{|c|}{}\\
{\tt IPROC} & {\tt ID}/{\tt IL} & {\tt TYPINT} & Process\\
\multicolumn{4}{|c|}{}\\
\hline\hline
\multicolumn{4}{|c|}{}\\
6000+{\tt ID} & {\tt ID}=1,...,6 & {\tt QED/QCD} & $\gamma\gamma/gg \rightarrow q\bar{q}$ (resp. d,u,s,c,b,t)\\
              & {\tt ID}=11  & {\tt QCD} & $gg \rightarrow q\bar{q}$ (all flavours)\\
              & {\tt ID}=13  & {\tt QCD} & $gg \rightarrow gg+q\bar{q}$ (all flavours)\\
              & {\tt ID}=7,8,9 & {\tt QED}     & $\gamma\gamma \rightarrow l^+l^-$ (resp. $e^+e^-$,$\mu^+\mu^-$,$\tau^+\tau^-$)\\
\multicolumn{4}{|c|}{}\\
9900+{\tt ID} & as for {\tt IPROC}=1600      & {\tt QED/QCD} & $\gamma\gamma/gg \rightarrow H$\\
\multicolumn{4}{|c|}{}\\
9800          & -            & {\tt QED/QCD} & $\gamma\gamma/gg \rightarrow \gamma\gamma$\\
\multicolumn{4}{|c|}{}\\
\hline
\end{tabular}
\end{center}
\caption{Special {\tt ID,TYPINT} settings for {\tt IPROC}=6000, and the new processes {\tt IPROC}=9800 and 9900; together 
with {\tt TYPEPR='EXC'}, these return the color-singlet $J_Z=0$ cross-sections.}
\label{t7}
\end{table}

\section{Structure of the program} \label{appa}

We give here a list of the {\tt FORTRAN} files that come with {\tt DPEMC}. They have been tested to comply with
{\tt HERWIG} version 6.5. The first set of files contain the original {\tt HERWIG} routines, which needed modifications for our purpose:
\begin{description}
\item [{\tt herwig6500.f}]: the original source code of {\tt HERWIG}; the routines {\tt HWSFUN}, {\tt HWEGAM}, {\tt HWEPRO} and {\tt HWHQPM}, 
superseded by the following files, are renamed to avoid conflicts;
\item [{\tt hwsfun65.f}]: structure function evaluation;
\item [{\tt hwegam65.f}]: chooses the flux and generates Pomeron or photon momentum fractions and virtualities; 
\item [{\tt hwepro65.f}]: decides on the hard process;
\item [{\tt hwhqpm65.f}]: calculates the cross-section for $gg \rightarrow q{\bar q}, \, gg$ in the $J_z = 0$ mode, 
for $\gamma\gamma \rightarrow q{\bar q}, \, W^+W^-, \, l^+l^-$ as in the original implementation of {\tt HERWIG}, and for 
$\gamma\gamma \rightarrow q{\bar q}$ in the $J_z = 0$ mode.
\end {description}

\noindent The following files contain the new routines needed to run {\tt DPEMC}:
\begin{description}
\item [{\tt flux65.f}]: implementation of the Pomeron and photon fluxes inside a proton or heavy ion;
\item [{\tt h1qcd.f}]: parton distribution functions inside the Pomeron;
\item [{\tt pomstr.f}]: user defined Pomeron structure function;
\item [{\tt hwhigp65.f}]: calculates the cross sections for $gg \rightarrow H$ and $\gamma\gamma \rightarrow H$;
\item [{\tt hwhqpp.f}]: evaluates the cross sections for $gg \rightarrow \gamma\gamma$ in the case of exclusive 
diphoton production and for $\gamma\gamma \rightarrow \gamma\gamma$ in the heavy-ion collisions;  
\item [{\tt hwnorm.f}]: calculates the normalization of parton distribution functions inside the Pomeron;
\item [{\tt hwfxer.f}]: corrects the event record.
\end{description}

\noindent In addition to the above mentioned files, a file {\tt vegas.f}, containing the program {\tt VEGAS} \cite{vegas} for numerical 
integration, is also needed to obtain the normalization of Pomeron parton distribution functions. The example main programs provided
with the package use {\tt HBOOK} for histogramming, and thus need to be linked against the CERN program library \cite{cernlib}.

An example main program is in {\tt example.f}, and produces 10000 exclusive $c\bar{c}$ events in the Bialas-Landshoff model. It is compiled by 
executing {\tt gmake -f makeExample}.

\section{} \label{appc}

\end{document}